\documentclass[12pt]{iopart}
\usepackage{graphicx}
\newcommand{\be}{\begin{equation}}
\newcommand{\ee}{\end{equation}}
\newcommand{\ba}{\begin{eqnarray}}
\newcommand{\ea}{\end{eqnarray}}
\newcommand{\baa}{\begin{eqnarray}}
\newcommand{\eaa}{\end{eqnarray}}
\newcommand{\ed}{\end{document}}
\newcommand{\lab}[1]{\label{#1}}
\newcommand{\re}[1]{(\ref{#1})}
\newcommand{\ci}[1]{\cite{#1}}
\renewcommand{\baselinestretch}{1.2}
\begin{document}

\title[Quantum particle under dynamical confinement]{Quantum particle under dynamical confinement:\\ From quantum Fermi acceleration to high harmonic generation}
\author{S.\ Rakhmanov$^{a,c}$, C.\ Trunk$^b$ and D.\ Matrasulov$^c$}
\address{$^a$Chirchik State Pedagogical University, 104 Amur Temur Str., Chirchik, 111700,Uzbekistan\\ $^b$  Technische Universit\"at Ilmenau, Postfach 100565, 98684 Ilmenau, Germany \\ $^c$ Turin Polytechnic University in Tashkent, 17 Niyazov Str., 100095,  Tashkent, Uzbekistan}

\begin{abstract}
Quantum dynamics of a particle confined in a box with time-dependent wall is revisited by considering some unexplored aspects of the problem. In particular,  the case of dynamical confinement in a time-dependent box in the presence of  purely time-varying external potential is treated by obtaining exact solution. Also, some external potentials approving separation of space and time variables in the  Schr\"odinger equation with time-dependent boundary conditions are classified. Time-dependence of the average kinetic energy and average quantum force are analyzed. A model for optical high harmonic generation in the presence of dynamical confinement and external linearly polarized monochromatic field is proposed.
\end{abstract}

\section*{Acknowledgement}
  \includegraphics[height=7.0mm]{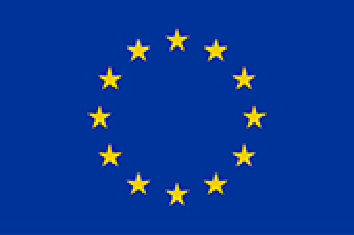} This research is partially supported by European Union’s Horizon 2020 research and innovation programme under the Marie Sklodowska-Curie grant agreement ID: 873071, project SOMPATY (Spectral Optimization: From Mathematics to Physics and Advanced Technology).

\section{Introduction}

Dynamical confinement in quantum mechanics attracted
 much attention during past few decades.
It is described in terms of the Schr\"odinger equation with time-dependent boundary conditions.  

Early treatments of the problem date back to Doescher, who explored basic aspects of the problem \cite{doescher}. Munier et al.\ considered more detailed treatment of the problem and computed some physically observable quantities for the problem of time-dependent box \cite{mun81}. Later Makowsky \cite{mak91}-\cite{mak923} and Razavy \cite{razavy01,razavy02} presented a systematic study of the problem, by considering one-dimensional box with moving walls and classifying time-dependence of the wall approving exact solution of the Schr\"{o}dinger equation with time-dependent boundary conditions.  Unitary transformation mapping time-dependent box to that with fixed walls was found in \cite{razavy01,razavy02} using an approach developed earlier by Berry and Klein \cite{berry}. Some aspects of the problem of quantum box with moving walls and its applications to dynamical Casimir effect was studied in a series of papers by Dodonov et al.\ \cite{dodonov01}-\cite{dodonov04}. Berry phase in time-dependent box was considered in \cite{per,Kwon,Wang}. Seba considered the problem of time-dependent box in the context of quantum Fermi acceleration \cite{seb90}. Application of the time-dependent box to the problem of confined quantum gas was considered in \cite{nakamura01,nakamura02}, where quantum force operator for dynamical confinement was introduced. 

The problem of hydrogen atom confined in time-dependent spherical box was considered in \cite{Our03}. Time-dependent harmonic oscillator which is directly related to time-dependent quantum box was presented in a series of papers by Lewis \cite{lewis01,lewis02}. Different aspects of the problem of dynamical confinement was studied in \cite{pinder} -\cite{glas09}. Inverse problem for dynamical confinement, i.e.\ the problem of recovering boundary's time-dependence from existing solution is considered in \ci{james}. Dynamical confinement in a half-line is studied in \ci{smith}. The probem of time-dependent Neumann boundary conditions is treated in \ci{duca}. Extension of the dynamical confinement to relativistic case   by considering Dirac equation for time-dependent box  was done in \cite{sobirov}. Time-dependent quantum graphs have been considered in the Refs.\cite{matrasulov,nikiforov,smilansky}.  

Despite the fact that considerable aspects of the problem of dynamical confinement have been considered, some issues in the topic are still remaining as less- or not studied. This concerns such aspects as time-dependent Neumann boundary conditions, non-adiabatic limit and exactly solvable models. Another important problem in this context is extension of the model to the case when time-dependent box interacts with an external potential. In such case, if the potential is position independent, the problem approves factorization of space and time variables.  

In this paper we address the problem of dynamical confinement in the presence of a external electromagnetic field. By assuming that time-dependence of the wall’s position approves separation of space and time-variables, we obtain general solution of the problem and compute such physically observable characteristics, as average kinetic energy and average force. Moreover, we consider the case of dynamical confinement driven by external linearly polarized monochromatic optical field. For this system, we study high harmonic generation induced by optical field. 

This paper is organized as follows.  In the next section we briefly recall the problem of time-dependent boundary conditions for the Schrodinger equation on a quantum box. In Section 3 we consider a particle in a time-dependent box in the presence of external potentials with the focus on
exactly solvable cases, i.e., when the problem allows factorization of the time- and space variables. Section 4 presents the treatment of the average kinetic energy of the particle and of the average quantum force acting to the particle by moving wall. Section 5 presents a quantum optics model for dynamical confinement
by considering high harmonic generation, Fermi acceleration and average quantum force as a function of time.

\section{Dynamical confinement in 1D quantum box} \label{Ilmenau}

Here, following \ci{mak91,razavy01}, we briefly recall the problem of time-dependent boundary conditions in quantum mechanics, by considering 1D quantum box with moving wall.
Consider a particle confined between two infinitely high walls. The position of the left wall is assumed to be fixed at $x=0$, while the right one  moves according to some positively determined function $L(t)$ which is a smooth function, $L:[0,\infty)\rightarrow(0,\infty)$. Then the particle dynamics in such a box is described in terms of the following  time-dependent Schr\"odinger equation ($\hbar=m=1$):

\be 
i\frac{\partial \Psi(x,t)}{\partial t}=H(x,t)\Psi(x,t),
\quad t\in [0,\infty), x\in [0, L(t)]
\lab{shr0001}
\ee
with the Hamiltonian

\be 
H(x,t)=-\frac{1}{2}\frac{\partial^2}{\partial x^2}+V(x,t).
\lab{ham1} 
\ee
Here, for simplicity, we assume the potential $V$ to be continuous 
on its domain $\{(x,t) \, | \, t\in [0,\infty), x\in [0, L(t)]\}$.
We impose, for Equation \re{shr0001}, the following Dirichlet boundary conditions given at the interval $[0, L(t)]$:

\begin{equation}
\Psi (x,t)|_{x=0}=\Psi (x,t)|_{x=L(t)}=0.
 \label{bc}
\end{equation}
Introducing a new coordinate

\be 
y=\frac{x}{L(t)}, \lab{trans001}
\ee
Equation \re{shr0001} can be rewritten as
\be
i\frac{\partial \Psi(y,t)}{\partial
t}=-\frac{1}{2L^2}\frac{\partial^2\Psi(y,t)}{\partial
y^2}+i\frac{\dot{L}}{L}y\frac{\partial\Psi(y,t)}{\partial y}+V(y
L(t),t)\Psi(y,t), \lab{shr001}
\ee
where $\dot{L}=dL/dt$ and new boundary conditions are given by

\be 
\lab{Oberhof}
\Psi(y,t)|_{y=0}=\Psi(y,t)|_{y=1}=0$ for all $t\in[0,\infty).
\ee

However, such transformation leads to breaking of the self-adjointness of the problem, i.e., the Schr\"odinger operator in the right hand side of  \re{shr001} is not self-adjoint. Therefore, one needs to recover self-adjointness using the transformation
\be
\Psi(y,t)=\sqrt{2/L}\exp{\Biggl( \frac{i}{2}L\dot{L} y^2
\Biggl)}\varphi(y,t), \lab{trans002}
\ee
that reduces Equation \re{shr001} to the following form \ci{mak91}
 \be
 i\frac{\partial\varphi(y,t)}{\partial
t}=-\frac{1}{2L^2}\frac{\partial^2 \varphi(y,t)}{\partial y^2}+
\Biggl( \frac{1}{2} L\ddot{L}y^2+V \Biggl)\varphi(y,t),
\lab{shr002}
\ee
where $\ddot{L}=d\dot{L}/dt$ and $\varphi(y,t)$ satisfies the boundary conditions  \re{Oberhof}. We mention that \re{shr002} can also be obtained from
Equation \re{shr0001} via some unitary transformation, cf.\ \ci{razavy02}.

In this section we consider the case $V=0$. Assume that the 
expression $4L^3\ddot{L}$ is a non-positive constant for all $t\in [0,\infty)$, i.e.\
\be 
0\leq B^2=-4L^3\ddot{L}=\textnormal{const} \lab{const001}
\ee
for some real $B$. Introduce a new “time variable” $\tau$ via
\be
\tau(t)=\int_{0}^{t}\frac{ds}{[L(s)]^2}.
\ee
Equation \re{shr002} reduces to 
\be 
i\frac{\partial\varphi(y,\tau)}{\partial
\tau}=-\frac{1}{2}\frac{\partial^2 \varphi(y,\tau)}{\partial
y^2}-\frac{1}{8}B^2y^2\varphi(y,\tau), \lab{shr003}
\ee
see also \ci{mak91}.
The solution of \re{shr003} can now be factorized as
\be
\varphi(y,\tau)=f(\tau)\Phi(y).
\ee
 Using a separation constant $K$,  the equation for $\Phi$ reduces to the Kummer equation
\be 
z\frac{d^2 U}{dz^2}+\Biggl( \frac{1}{2}-z \Biggl)\frac{d U}{d z}+\frac{1}{4}(\kappa^2-1)U=0, \lab{eq001}
\ee
where $z=(iB/2)y^2, \kappa^2=4K/iB, U(z)=\exp{(z/2)}\Phi(z)$. Hence (see 13.1.13 from \ci{Abramowitz}) the Kummer function $z^{1/2}M\Big(\frac{3-\kappa^2}{4},\frac{3}{2},z\Big)$ is a solution of Equation \re{eq001}. Hence, a solution for
\be \lab{Elgersburg}
H_0\Phi(y)=K \Phi(y)
\quad \mbox{with }H_0:=-\frac{1}{2}\frac{d^2 }{d y^2}- \frac{1}{8} B^2y^2.
\ee
 is of the form
\be
\Phi(y)=Cy M \Biggl(\frac{3iB-4K}{4i B},\frac{3}{2},\frac{iB}{2}y^2\Biggl)e^{-\frac{iB}{4}y^2}, \lab{unsysphi}
\ee
where $C$ is the normalization constant. 
Exact  solution of Equation \re{shr0001} can  be written as \ci{mak91}
\be
\Psi(x,t) = \frac{Cx}{\sqrt{L^{3}}}
e^{\frac{i}{2} x^2 \left(
\frac{\dot{L}}{L}-\frac{B}{2L^2}\right)-i K \tau(t)} 
 M \left(\frac{3i B-4K}{4i B},\frac{3}{2},\frac{iB}{2}\frac{x^2}{L^2}\right). \lab{s001} 
\ee
Note that $\Psi(0,t)=0$ for all $t$. However, $\Psi(L,t)=0$ if and if 
\be
M\left(\frac{3iB-4K}{4iB},\frac{3}{2},\frac{iB}{2}\right)=0. \lab{kumm001}
\ee
Therefore the boundary condition \re{bc} is satisfied if and only if $K$ equals a zero of the Kummer function in \re{kumm001}. Denote these zeros by $K_n$, $n\in N$. Define the function $\Psi_n$ via the right hand side of Equation \re{s001} where $K$ is replaced by $K_n$.

It is important to mention that the time-dependent Dirichlet boundary conditions imposed for \re{shr0001} approve norm conservation. Indeed,
 let $N(t)$ be the norm at time $t$
 as the $L^2$-norm of $\Psi$ with respect to the spatial variable $x$,
\be
N(t):= ||\Psi(x,t)||^2=\int^{L(t)}_{0}|\Psi|^2dx.
\ee
Then, for the time-derivative we have
\be
\frac{dN}{dt}(t)=\int^{L(t)}_{0}\frac{\partial}{\partial
t}|\Psi(x,t)|^2dx+\dot{L}(t)|\Psi|^2|_{x=L(t)} =
\int^{L(t)}_{0}\frac{\partial}{\partial
t}|\Psi(x,t)|^2dx
\lab{eq3}
\ee
Taking into account \re{shr0001}, \re{bc} and
\be
 i\int^{L(t)}_{0}\frac{\partial}{\partial
 t}|\Psi|^2dx=
 \frac{1}{2}\left.\left(
 \Psi\frac{\partial\Psi^{*}}{\partial x}-\Psi^{*}\frac{\partial\Psi}{\partial
 x}
 \right) 
 \right|^{x=L(t)}_{x=0}=0,
\lab{eq7}
\ee
we have norm conservation $\frac{dN}{dt}(t)=0$.

\section{Dynamical confinement in the presence of  external potential:\\ Exactly solvable models}

The model in Section \ref{Ilmenau} approves factorization of the variables in the case of constraint \re{const001}. However, when the system interacts with an external position-independent time-varying field, factorization is also possible. Here we consider time-dependent quantum box driven by external purely time-dependent  potential $V$. The dynamics of the system is governed by the following time-dependent Schr\"odinger equation:
\begin{equation}
i  \frac {\partial \Psi(x,t)}{\partial t}=-\frac{1}{2} \frac{\partial^2
\Psi(x,t)}{\partial x^2} +V(t)\Psi(x,t)\label{shred01}
\end{equation}
The boundary conditions for this equation are imposed as in \re{bc}. Following the previous section, we  transform the boundary conditions into time-independent ones by introducing a new coordinate $y$ which is given by \re{trans001}. Using the transformation of the wave function given by Equation \re{trans002} and
under the assumption that $L(t)$ fulfills the condition in Equation \re{const001}, 
we have with Equation \re{shr002}
\begin{equation}
     iL^2\frac{\partial\varphi(y,t)}{\partial
t}=-\frac{1}{2}\frac{\partial^2 \varphi(y,t)}{\partial y^2}-
 \frac{1}{8} B^2y^2\varphi(y,t)+L^2V(t) \varphi(y,t),
 \lab{tdbc001}
\end{equation}

Let $\Phi_n$  satisfy \re{Elgersburg} with Dirichlet
boundary conditions \re{kumm001}
with respect to the eigenvalue $K=K_n$ and denote its normalization constant by
$C_n$. It is well-known 
that the system $\{\Phi_n\}_n$ forms an orthonormal basis of $L^2(0,1)$.
We choose as an Ansatz for the solution of \re{tdbc001}

\begin{equation}
    \varphi(y,t)=\sum_nC_n(t)\Phi_n(y) \lab{expand}
\end{equation} 
By substituting \re{expand} into Equation
(\ref{tdbc001}), we (formally) obtain
\begin{equation}
iL^2\sum_n\dot{C}_n(t)\Phi_n(y)=\sum_nC_n(t)H_0\Phi_n(y)+L^2V(t)\sum_nC_n(t)\Phi_n(y).
\end{equation}

Then, after multiplying both sides of equation to $\Phi_m^*(y)$, integrating over $y$ and using the orthonormal condition for a basis, $\int_0^1\Phi_m^*(y)\Phi_n(y)dy=\delta_{mn}$, we have
\begin{equation}
     iL^2\dot{C}_n(t)=C_n(t)K_n+L^2V(t)C_n(t)
\end{equation}
It's solution is in the form
\begin{equation}
     C_n(t)=C_{n}(0)e^{-i\int_0^t\big(\frac{K_n}{L^2}+V(s)\big)ds}
     \lab{C_nt}
\end{equation} 
where $C_n(0)$ can be determined from a smooth initial
condition which insures $\sum_n|C_n(0)|^2<\infty$.
Finally, one obtains the general solution for Equation (\ref{shred01})
\begin{eqnarray}
     \Psi(x,t)=\sum_nC_n(0)e^{-i\int_0^t\big(\frac{K_n}{L^2}+V(s)\big)ds}
		C_n\sqrt{\frac{2}{L^3}}xM \Biggl(\frac{3iB-4K_n}{4i B},\frac{3}{2},\frac{iB}{2}\frac{x^2}{L^2}\Biggl) \nonumber \\
     \times e^{-\frac{iB}{4}\frac{x^2}{L^2}+\frac{i}{2}\frac{\dot{L}}{L}x^2}.
\end{eqnarray}

\medskip
It is worthful  to consider  some other potentials approving factorization of variables in the   Schr\"{o}dinger equation with time-dependent boundary conditions. One of them is interaction proportional to inverse square of the distance given as
 $$V=\frac{\alpha}{x^2}$$ 
 where $\alpha$ is constant. For this case Equation \re{shr002} can be written as
 \be
 i\frac{\partial\varphi(y,t)}{\partial
t}=-\frac{1}{2L^2}\frac{\partial^2 \varphi(y,t)}{\partial y^2}+
\Biggl( \frac{1}{2} L\ddot{L}y^2+\frac{\alpha}{L^2y^2} \Biggl)\varphi(y,t),
\ee
Variables of this equation can be separated, provided $L(t)$ fulfills \re{const001}.

Similarly to the above, one can show that potential in the form $V=x\varepsilon(t)$ also approves of separation time and space variables and one obtains Schr\"{o}dinger equation for anharmonic oscillator given as
\be \lab{Bischleben}
 i\frac{\partial\varphi(y,t)}{\partial
t}=-\frac{1}{2L^2}\frac{\partial^2 \varphi(y,t)}{\partial y^2}+
\Biggl( \frac{1}{2} L\ddot{L}y^2+Ly\varepsilon(t) \Biggl)\varphi(y,t),
\ee
Variables of this equation can be separated when $L^3\ddot{L} = \textnormal{const} = \beta$ and $L^3\varepsilon(t) = \textnormal{const} = \gamma$ conditions are fulfilled. From those two conditions one can see $\varepsilon(t)=\frac{\gamma}{\beta}\ddot{L}$
and \re{Bischleben} becomes
 \be
 iL^2\frac{\partial\varphi(y,t)}{\partial
t}=-\frac{1}{2}\frac{\partial^2 \varphi(y,t)}{\partial y^2}+
\frac{1}{2} \beta y^2 \varphi(y,t)+\gamma y \varphi(y,t).
\ee

Finally, for potential given in the form of nonlinearly polarized monochromatic field given by  $V=x^2\epsilon\cos\omega t,$ where $\epsilon$, $\omega$ are strength and frequency of external field, one can also factorize space and time variables and have

\be
 i\frac{\partial\varphi(y,t)}{\partial
t}=-\frac{1}{2L^2}\frac{\partial^2 \varphi(y,t)}{\partial y^2}+
\Biggl( \frac{1}{2} L\ddot{L}y^2+L^2y^2\epsilon\cos\omega t \Biggl)\varphi(y,t),
\ee
Conditions for factorization of variables for this equation given in the form of constraint $L^3\ddot{L}+2\epsilon L^4\cos\omega t = \textnormal{const} = \beta$.

\section{Average kinetic energy and quantum force induced by dynamical confinement}

Having found the solution of the Schr\"{o}dinger equation for time-dependent quantum box, one can compute physically observable variables, such as average kinetic energy and average (quantum) force. 
The average kinetic energy is determined as the expectation value of the kinetic energy operator:
\be 
\hat{H}=-\frac{1}{2}\frac{\partial^2}{\partial x^2}. \lab{hamil} 
\ee
The expectation value of energy is  given by
\be 
<E(t)> = \langle \Psi| \hat{H} | \Psi \rangle, 
\ee
where $|\Psi\rangle$ is a solution of \re{shred01}. We have

\be 
<E_k(t)> =\int_0^{L(t)}\Psi^*(x,t)
\Biggl(-\frac{1}{2}\frac{\partial^2}{\partial x^2} \Biggl)
\Psi(x,t)dx=\frac{1}{2}\int_0^{L(t)} \Biggl| \frac{\partial
\Psi(x,t)}{\partial x}\Biggl|^2 dx   \lab{avkin} 
\ee 
or with \re{trans002}
$$<E_k(t)>=\frac{1}{L^2}\int_0^1\Biggl|\frac{\partial\varphi(y,t)}{\partial y}\Biggl|^2dy + \frac{2\dot{L}}{L}\textnormal{Im}\Biggl(\int_0^1y\varphi^*(y,t)\frac{\partial\varphi(y,t)}{\partial y}dy\Biggl)+$$
$$\dot{L}^2\int_0^1y^2|\varphi(y,t)|^2dy =\frac{1}{L^2}S_0+ \frac{2\dot{L}}{L}\textnormal{Im}(S_1)+\dot{L}^2S_2.$$
Explicit expressions for  $S_0,\;$ $S_1\;$ and $S_2\;$ are provided in Appendix A.

Quantum force can be determined as the expectation value of the force operator as
\be 
\hat F=-\frac{\partial  \hat{H}}{\partial L(t)}   . \lab{for}
\ee

Then for the expectation value of the force operator one has \ci{nakamura01}
\be  
<F(t)>=-\frac{\partial \langle E_k(t)
\rangle}{\partial L}= \frac{2}{L^3}S_0+ \frac{2\dot{L}}{L^2}\textnormal{Im}(S_1).
\lab{opfor2}
\ee

\section{Quantum Fermi acceleration and high harmonic generation in driven time-dependent box}

An important effect that can be realized in a quantum box with oscillating walls is the so-called Fermi acceleration in quantum regime, or quantum Fermi acceleration. It is a quantum analog of the classical Fermi acceleration that occurs in bouncing balls colliding with oscillating wall. In classical regime, unbounded growth of the average kinetic energy of a particle can be observed in such a system. Quantum Fermi acceleration in a time-dependent box was studied in \ci{razavy02,seb90}. Here we extend these studies to the case of interaction (in addition to the interaction with oscillating wall) with an external time-periodic potential.

\begin{figure}[t!]
\centering
\includegraphics[totalheight=0.4\textheight]{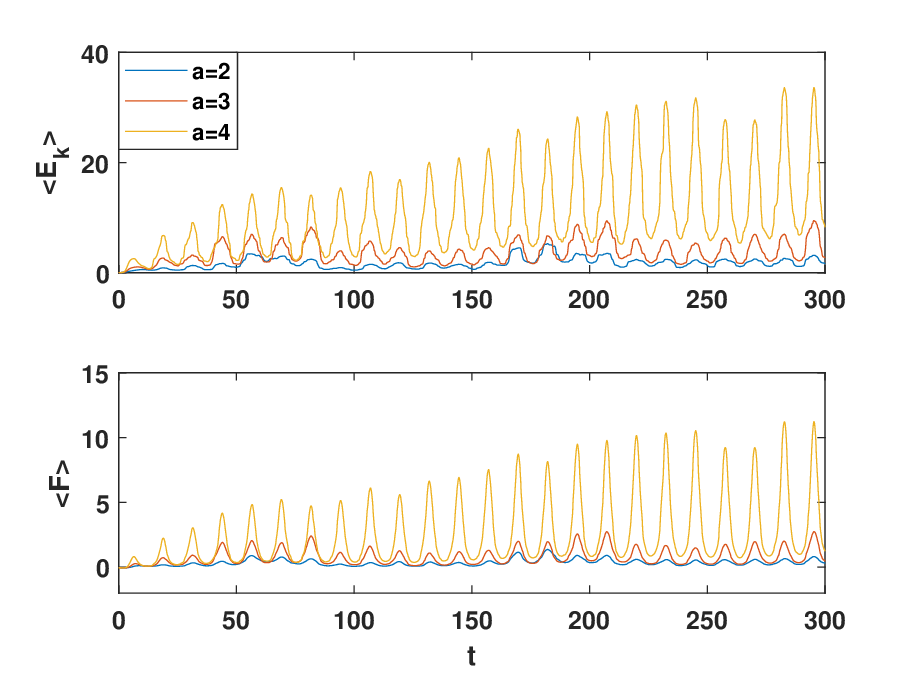}
 \caption{Average kinetic energy and force as a function of time at different values of amplitude of oscillating wall $a$ for $L_0=10$, $\omega_0=0.5$, $\epsilon=0.1$ and $\omega=0.05$ ($L=L_0+a\cos\omega_0t$ and V=$\epsilon x\cos\omega t$).} \label{fig:1}
\end{figure}

We consider a version of the dynamically confined system which can be experimentally realized in optics. Namely, we propose a model for time-dependent box driven by linearly polarized monochromatic field given by 
\be
V(x,t) =\epsilon x\cos\omega t,
\lab{field1}
\ee
where $\epsilon$ and $\omega$ are the field strength and the frequency,  respectively.

In the following we consider a quantum particle in a box with oscillating wall and interacting with the external linearly polarized monochromatic field given by \re{field1}. Wall's oscillation is assumed to be given as 
\begin{equation*}
L=L_0+a\cos\omega_0t.
\end{equation*}

In this case, Equation \re{shr002}
cannot be separated and one has to solve it numerically. 
Here we will use the following Ansatz for  $\varphi(y,t)$:
\begin{equation}\large
\varphi (y,t)=\sum_n C_n(t) \sin \pi ny, \label{func}
\end{equation}
where $\phi_n (y):= \sin \pi ny$ solves the equation
$
-\frac{1}{2} \frac {d^2\phi _n}{dy^2} =\frac{\pi^2n^2}{2} \phi _n
$
with Dirichlet boundary conditions on the interval $[0,1]$.

\begin{figure}[t!]
\centering
\includegraphics[totalheight=0.4\textheight]{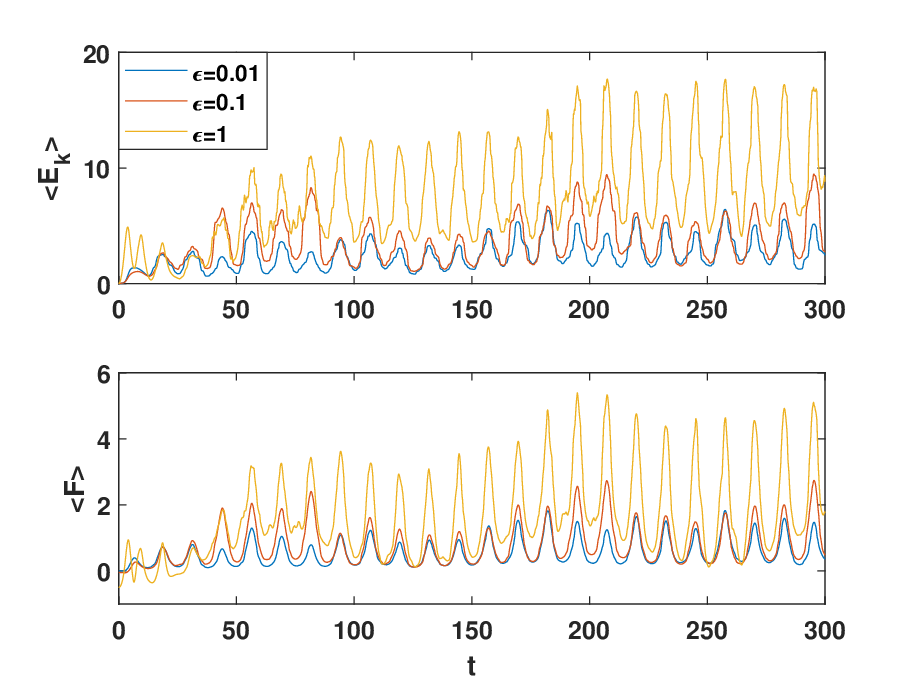}
 \caption{Average kinetic energy and force as a function of time at different values of field strength $\epsilon$ for $L_0=10$, $a=3$, $\omega_0=0.5$ and $\omega=0.05$ ($L=L_0+a\cos\omega_0t$ and V=$\epsilon x\cos\omega t$).} \label{fig:2}
\end{figure}

For the coefficients $C_n$ we have the following system of differential equations
\begin{equation}\large
i L^2\dot {C}_n(t) = C_n(t) \frac{\pi^2n^2}{2} +\sum_m V_{nm}C_m(t)
 \label{eq2}
\end{equation}
where
$V_{nm}=L^3\ddot{L}I_{1nm}+2\epsilon L^3\cos\omega tI_{2nm}$
and
$$ 
I_{1nm}=\int_0^1y^2\phi_n^*\phi_mdy=  \left\{\begin{array}{l} \quad\qquad\frac{1}{6}+\frac{1}{4n^2\pi^2},\qquad\qquad \; n=m \\  \frac{1}{\pi^2}\Big( \frac{(-1)^{m-n}}{(m-n)^2}-\frac{(-1)^{m+n}}{(m+n)^2} \Big),  \qquad  n\neq m \end{array}\right. 
$$
and
$$
I_{2nm}=\int_0^1y\phi_n^*\phi_mdy=\left\{\begin{array}{l} \quad\qquad\qquad\frac{1}{4},\qquad\qquad\qquad \quad \; \; n=m \\  \frac{1}{2\pi^2}\Big( \frac{(-1)^{m-n}-1}{(m-n)^2}-\frac{(-1)^{m+n}-1}{(m+n)^2} \Big),  \qquad  n\neq m \end{array}\right.
$$
We solve \re{eq2} numerically by choosing initial condition as $C_1(0) =1$ and
 $C_n(0)=0$ for $n\neq 1$. Having found $C_n(t)$ one constructs $\phi_n$ and $\Psi$
 via \re{trans001} and \re{trans002}.

In Figure \ref{fig:1} time-dependence of the average kinetic energy of the particle $<E_k>$ and the quantum force $<F>$ acting on the wall are plotted at different values of the wall's oscillation amplitude $a$. For smaller values of $a$ both average kinetic energy and force are periodic, while at higher values they become quasi periodic and certain growth of the "peaks" can be observed. However, suppression of the growth occurs as time elapses. 
 
\begin{figure}[t!]
\centering
\includegraphics[totalheight=0.4\textheight]{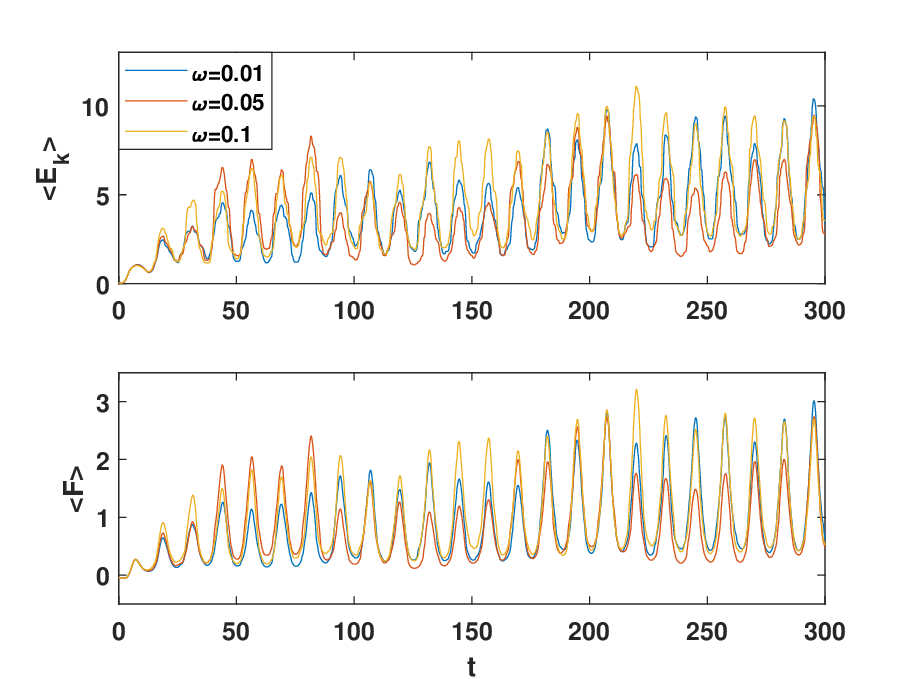}
 \caption{Average kinetic energy and force as a function of time at different values of frequency of external field $\omega$ for $L_0=10$, $a=3$, $\omega_0=0.5$ and $\epsilon=0.1$ ($L=L_0+a\cos\omega_0t$ and V=$\epsilon x\cos\omega t$).} \label{fig:3}
\end{figure}

 Figure \ref{fig:2} presents the time-dependence of the average kinetic energy and the force at different values of the external field strength $\epsilon$.
 The behavior of $<E_k(t)>$ and  $<F(t)>$ are similar to that in Figure \ref{fig:1}, which implies similarity of the roles of  $\epsilon$ and $a$  in the particle dynamics. In other words, particle "feels" both, oscillating wall and external monochromatic field as a periodic perturbation. 

\begin{figure}[t!]
\centering
\includegraphics[totalheight=0.4\textheight]{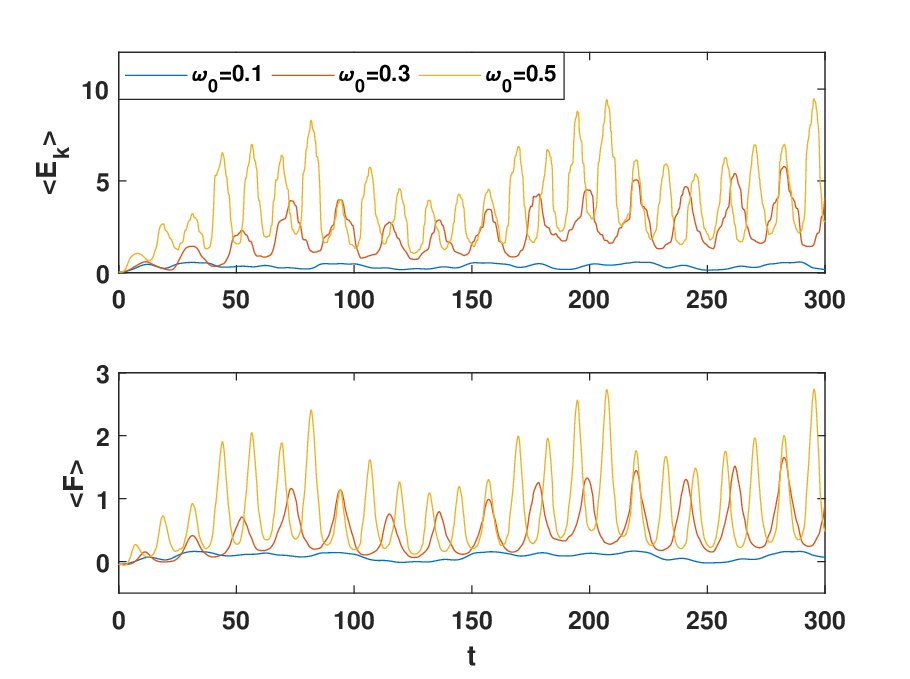}
 \caption{Average kinetic energy and force as a function of time at different values of frequency of oscillating wall $\omega_0$ for $L_0=10$, $a=3$, $\epsilon=0.1$ and $\omega=0.05$ ($L=L_0+a\cos\omega_0t$ and V=$\epsilon x\cos\omega t$).} \label{fig:4}
\end{figure}

In Figures \ref{fig:3} and \ref{fig:4}, $<E_k(t)>$ and  $<F(t)>$ are plotted at different values of the external field and oscillating boundary frequency for $L_0=10$, respectively. Qualitatively plots look similar to those in Figures \ref{fig:1}  and \ref{fig:2}. 

 
An important effect which can be considered in the system time-dependent quantum box+ external monochromatic field is optical high harmonic generation (HHG) induced by interaction of the time-dependent box with external optical field given by \re{field1}. Evolution of the whole  system, "dynamical box + optical field" is governed by \re{shr0001}. 

Detailed description of the
high harmonic generation in quantum regime can be found in
\ci{Boydbook}. Here we will focus on the role of confinement in
optical harmonic generation.  The main physical characteristics of
such process is the average dipole moment which is given by
\ci{Boydbook}
$$
<d(t)> = -<\Psi(x,t)|x|\Psi(x,t)>.
$$

The spectrum of high harmonic generation (HHG) is characterized by the quantity \ci{Boydbook}
\be 
I(\nu)= |< d(\nu)>|^2=\Biggl|\frac{1}{T}\int_{0}^{T} e^{-i\nu t}< d(t)>dt\Biggl|^2,
\lab{spectr1}
\ee 
where $T$ is the total duration of interaction.

\begin{figure}[t!]
\centering
\includegraphics[totalheight=0.38\textheight]{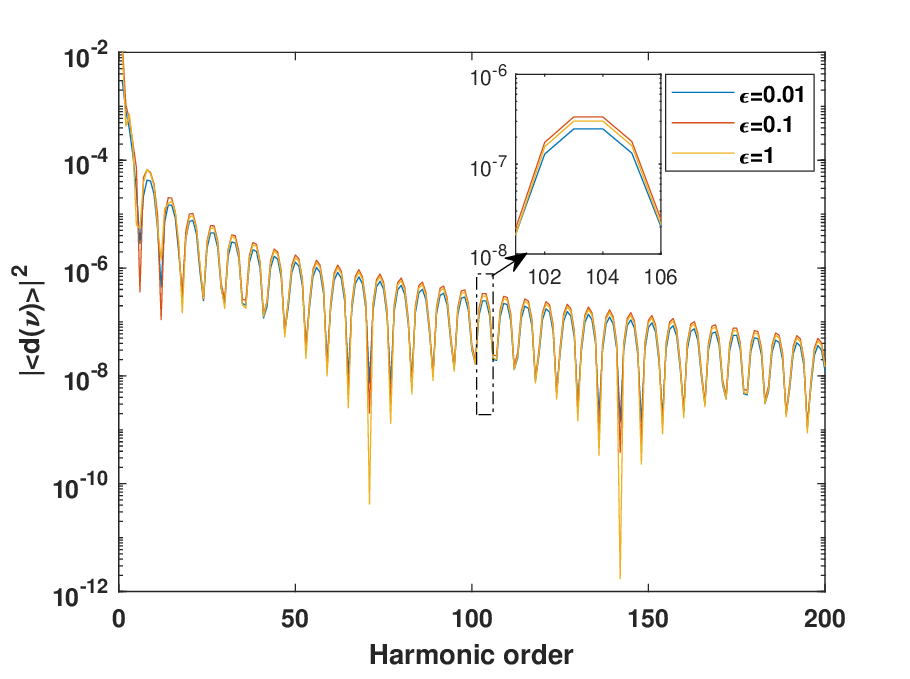}
 \caption{Spectrum of HHG as a function of harmonic order at different values of field strength $\epsilon$ for $L_0=10$, $a=3$, $\omega_0=0.5$, $\omega=1$ and T=200 ($L=L_0+a\cos\omega_0t$ and $V=\epsilon x\cos\omega t$).} \label{fig:8}
\end{figure}

\begin{figure}[t!]
\centering
\includegraphics[totalheight=0.38\textheight]{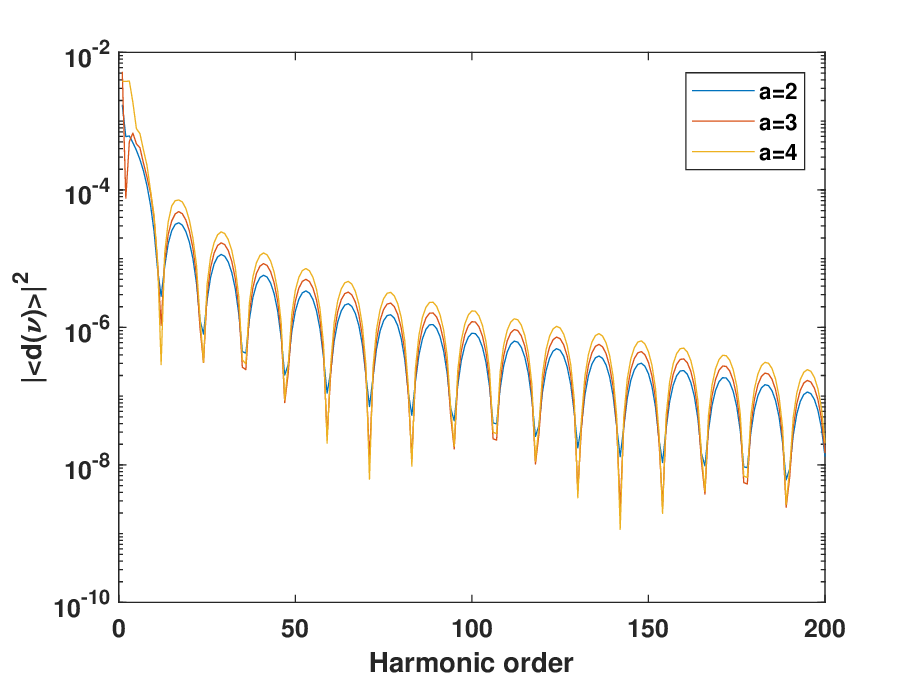}
 \caption{Spectrum of HHG as a function of harmonic order at different values of amplitude of oscillating box $a$  for $L_0=10$ $\epsilon=0.1$, $\omega_0=0.5$, $\omega=1$ and $T=100$ ($L=L_0+a\cos\omega_0t$ and $V=\epsilon x\cos\omega t$).} \label{fig:9}
\end{figure}

Figure \ref{fig:8} shows plots of the spectrum of harmonic generation as a function of harmonic order at different values of  external field strength for $L_0=10$, $a=3$, $\omega_0=0.5$, $\omega=1$ and $T=200$. The plot shows that  the HHG intensity strongly depends on the  field strength. For higher values of $\epsilon$, one can observe  increasing of intensity. Figure
\ref{fig:9}  presents plots of HHG-spectra at different amplitudes of oscillating box. An important feature of the HHG-spectra presented in Figures \ref{fig:8}  and \ref{fig:9}  is the existence of a plateau in the curves, i.e.\ there is rather side range of frequencies having the same intensity of generation (emission). Such a feature is of importance from the viewpoint of applications of the model in attosecond physics, where wide range of generated high frequencies with high enough intensity is required.

\section{Conclusion}

In this paper we considered the problem of dynamical confinement in a time-dependent 1D quantum box interacting with external potentials. The main focus is given to find an exact solution of the problem. In particular, an exact solution of the problem  is obtained for the purely time-dependent external potential. Some other external potentials approving  factorization of  space and time variables in the time-dependent  Schr\"odinger equation with moving boundary conditions are classified.

Average kinetic energy and quantum force for the particle simultaneously subjected to the influence of dynamical confinement and external time-periodic field are analyzed as a function of time. A model for high harmonic generation in time-dependent box that can be experimentally realized in quantum optics, is proposed. The spectrum of high harmonics generated by such system is computed.

The proposed model  can be directly applied to the problems of  tunable quantum Fermi acceleration and quantum transport in low-dimensional confined systems arising in optics and condensed matter. An extension of the model to the case of 2D and 3D systems, where the role of boundaries geometry is very challenging is a task for forthcoming studies. 

\section*{Appendix A}

Here we will give explicit forms of the  quantities $S_0,\;$ $S_1\;$ and $S_2\;$
Using the solution of Equation \re{tdbc001} one can find $S_0,\;$ $S_1\;$ and $S_2\;$ that are defined as

$$S_0=\sum_{n,m}C_n^*(t)C_m(t)\Biggl[I_1+\frac{W_m}{6}I_2+\frac{W_n^*}{6}I_2^*+\frac{W_n^*W_m}{36}I_4-\frac{iBW_n^*}{12}I_5+\frac{iBW_m}{12}I_5^*+\frac{B^2}{4}I_6  \Big] $$

$$S_1=\sum_{n,m}C_n^*(t)C_m(t) \Biggl[ I_3 +\frac{W_m}{6}I_5^* -\frac{iB}{2}I_6 \Big]$$

$$S_2=\sum_{n,m}C_n^*(t)C_m(t)I_6$$

with $W_n=3iB-4K_n$ and $I_1,\;$ $I_2,\;$  $I_3,\;$ $I_4,\;$ $I_5,\;$ $I_6\;$ given by

$$I_1=\int_0^1M^*\Biggl(\frac{3i B-4K_n}{4i B},\frac{3}{2},\frac{iB}{2}y^2\Biggl)M\Biggl(\frac{3i B-4K_m}{4i B},\frac{3}{2},\frac{iB}{2}y^2\Biggl)dy$$

$$I_2=\int_0^1y^2M^*\Biggl(\frac{3i B-4K_n}{4i B},\frac{3}{2},\frac{iB}{2}y^2\Biggl)M\Biggl(\frac{7i B-4K_m}{4i B},\frac{5}{2},\frac{iB}{2}y^2\Biggl)dy$$

$$I_3=\int_0^1y^2M^*\Biggl(\frac{3i B-4K_n}{4i B},\frac{3}{2},\frac{iB}{2}y^2\Biggl)M\Biggl(\frac{3i B-4K_m}{4i B},\frac{3}{2},\frac{iB}{2}y^2\Biggl)dy$$

$$I_4=\int_0^1y^4M^*\Biggl(\frac{7i B-4K_n}{4i B},\frac{5}{2},\frac{iB}{2}y^2\Biggl)M\Biggl(\frac{7i B-4K_m}{4i B},\frac{5}{2},\frac{iB}{2}y^2\Biggl)dy$$

$$I_5=\int_0^1y^4M^*\Biggl(\frac{7i B-4K_n}{4i B},\frac{5}{2},\frac{iB}{2}y^2\Biggl)M\Biggl(\frac{3i B-4K_m}{4i B},\frac{3}{2},\frac{iB}{2}y^2\Biggl)dy$$

$$I_6=\int_0^1y^4M^*\Biggl(\frac{3i B-4K_n}{4i B},\frac{3}{2},\frac{iB}{2}y^2\Biggl)M\Biggl(\frac{3i B-4K_m}{4i B},\frac{3}{2},\frac{iB}{2}y^2\Biggl)dy$$

\end{document}